\documentclass{iopart}
\usepackage{amssymb}
\usepackage{graphicx}
\usepackage{graphics}
\usepackage{eepic,epsfig}

\begin{document}

\title[Vacuum densities for a thick brane in AdS spacetime]{Vacuum densities
for a thick brane \\
in AdS spacetime}
\author{A A Saharian and A L Mkhitaryan }

\address{Department of Physics, Yerevan State University\\
1 Alex Manoogian Street, 375025 Yerevan, Armenia}
\ead{saharian@ictp.it}

\begin{abstract}
For a massive scalar field with general curvature coupling parameter we
evaluate Wightman function, vacuum expectation values of the field square
and the energy-momentum tensor induced by a $Z_{2}$-symmetric brane with
finite thickness located on $(D+1)$-dimensional AdS bulk. For the general
case of static plane symmetric interior structure the expectation values in
the region outside the brane are presented as the sum of free AdS and brane
induced parts. For a conformally coupled massless scalar the brane induced
part in the vacuum energy-momentum tensor vanishes. In the limit of strong
gravitational fields the brane induced parts are exponentially suppressed
for points not too close to the brane boundary. As an application of general
results a special model is considered in which the geometry inside the brane
is a slice of the Minkowski spacetime orbifolded along the direction
perpendicular to the brane. For this model the Wightman function, vacuum
expectation values of the field square and the energy-momentum tensor inside
the brane are evaluated. It is shown that for both minimally and conformally
coupled scalar fields the interior vacuum forces acting on the brane
boundaries tend to decrease the brane thickness.
\end{abstract}


\section{Introduction}

\label{sec:introd}

Braneworlds naturally appear in string/M-theory context and provide a novel
setting for discussing phenomenological and cosmological issues related to
extra dimensions. Motivated by the problems of the radion stabilization and
the generation of cosmological constant, the role of quantum effects in
braneworlds has attracted great deal of attention \cite{Fabi00}-\cite{Mina07}%
. A class of higher dimensional models with compact internal spaces is
considered in \cite{Flac03b}. Many of treatments of quantum fields in
braneworlds deal mainly with the case of the idealized brane with zero
thickness. This simplification suffers from the disatvantage that the
curvature tensor is singular at the brane location. In addition, the vacuum
expectation values of the local physical observables diverge on the brane.
From a more realistic point of view we expect that the branes have a finite
thickness and the thickness can act as natural regulator for surface
divergences. The finite core effects also lead to the modification of the
Friedmann equation describing the cosmological evolution inside the brane.
In sting theory there exists the minimum length scale and we cannot neglect
the thickness of the corresponding branes at the string scale. The branes
modelled by field theoretical domain walls have a characteristic thickness
determined by the energy scale where the symmetry of the system is
spontaneously broken. Various models are considered for a thick brane.
Mainly, these models are constructed as solutions to the coupled
Einstein-scalar equations by choosing a suitable potential for the scalar
field. Vacuum fluctuations for a thick de Sitter brane supported by a bulk
scalar field with an axion like potential and the self-consistency of this
braneworld are investigated in \cite{Mina06a}.

In the present paper based on \cite{Saha07} we describe the effects of core
on properties of the quantum vacuum for a general plane symmetric static
model of the brane with finite thickness. The most important quantities
characterizing these properties are the vacuum expectation values of the
field square and the energy-momentum tensor. Though the corresponding
operators are local, due to the global nature of the vacuum, the vacuum
expectation values describe the global properties of the bulk and carry an
important information about the internal structure of the brane. As the
first step for the investigation of vacuum densities we evaluate the
positive frequency Wightman function for a massive scalar field with general
curvature coupling parameter. This function gives comprehensive insight into
vacuum fluctuations and determines the response of a particle detector of
the Unruh-DeWitt type moving in the brane bulk. The problem under
consideration is also of separate interest as an example with gravitational
and boundary-induced polarizations of the vacuum, where all calculations can
be performed in a closed form. The corresponding results specify the
conditions under which we can ignore the details of the interior structure
and approximate the effect of the brane by the idealized model. In addition,
as it will be shown below, the phenomenological parameters in the
zero-thickness brane models such as brane mass terms for scalar fields are
calculable in terms of the inner structure of the brane within the framework
of the model considered in the present paper.

The paper is organized as follows. In Section \ref{sec:WF} we consider the
Wightman function in the exterior of the brane for the general structure of
the core with Poincare invariance along the directions parallel to the
brane. By using the formula for the Wightman function, in Section \ref%
{sec:Outside} we investigate the vacuum expectation values of the field
square and the energy-momentum tensor. As an illustration of the general
results, in Section \ref{sec:flowerpot} we consider a model with Minkowskian
geometry inside the brane. For this model the vacuum expectation values
inside the core are investigated as well. The last section contains a
summary of the work.

\section{Wightman function}

\label{sec:WF}

\begin{figure}[tbph]
\begin{center}
\epsfig{figure=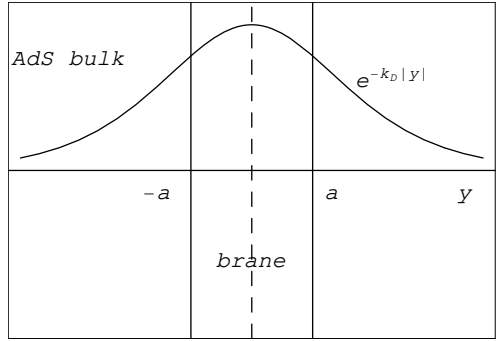,width=6.5cm,height=4.5cm}
\end{center}
\caption{The geometry of a thick brane on AdS bulk.}
\label{figbrane}
\end{figure}

We consider a brane with finite thickness $2a$ on background of $(D+1)$%
-dimensional AdS spacetime with the curvature radius $1/k_{D}$ (see figure %
\ref{figbrane}). As in the Randall-Sundrum (RS) 1-brane scenario
\cite{Rand99b} we assume that the model is $Z_{2}$-symmetric with
respect to the plane $y=0$ located at the brane center. The
spacetime is described by two distinct metric tensors in the
regions outside and inside the brane. The corresponding line
element has the form
\begin{equation}
ds^{2}=\left\{
\begin{array}{ccc}
e^{-2k_{D}|y|}\left( dt^{2}-d\mathbf{x}^{2}\right) -dy^{2}, & \mathrm{if} &
|y|>a, \\
e^{2u(y)}(dt^{2}-d\mathbf{x}^{2})-e^{2w(y)}dy^{2}, & \mathrm{if} & |y|<a,%
\end{array}%
\right. \;  \label{metric}
\end{equation}%
where $\mathbf{x}=(x^{1},\ldots ,x^{D-1})$ are the coordinates parallel to
the brane. We assume that the geometry inside the brane is Poincare
invariant along these directions. Due to the $Z_{2}$-symmetry the functions $%
u(y)$, $w(y)$ are even functions of $y$. These functions are continuous at
the core boundary: $u(a)=-k_{D}a,\;w(a)=0$. Here we assume that an
additional infinitely thin plane shell located at $|y|=a$ is present with
the surface energy-momentum tensor $\tau _{i}^{k}$, $\tau _{D}^{D}=0$. From
the Israel matching conditions one has%
\begin{equation}
u^{\prime }(a-)=-k_{D}+8\pi G\tau _{0}^{0}/(D-1),\;\tau _{i}^{k}=\tau
_{0}^{0}\delta _{i}^{k},\;i=1,2,\ldots ,D-1,  \label{matchcond2}
\end{equation}%
where $G$ is the Newton gravitational constant.

We are interested in the vacuum polarization effects for a scalar field with
general curvature coupling parameter $\xi $ propagating in the bulk
described by line element (\ref{metric}). The corresponding field equation
has the form%
\begin{equation}
\left( \nabla _{i}\nabla ^{i}+m^{2}+\xi R\right) \varphi =0,  \label{fieldeq}
\end{equation}%
where $R$ is the Ricci scalar for the background spacetime. As a first stage
for the evaluation of the vacuum expectation values (VEVs) for the field
square and the energy-momentum tensor (EMT) we consider the positive
frequency Wightman function. This function can be evaluated by using the
mode sum formula
\begin{equation}
\langle 0|\varphi (x)\varphi (x^{\prime })|0\rangle =\sum_{\alpha }\varphi
_{\alpha }(x)\varphi _{\alpha }^{\ast }(x^{\prime }),  \label{mfieldmodesum}
\end{equation}%
where $\left\{ \varphi _{\alpha }(x),\varphi _{\alpha }^{\ast }(x^{\prime
})\right\} $ is a complete orthonormalized set of positive and negative
frequency solutions to the field equation specified by the collective index $%
\alpha $.

The eigenfunctions can be presented in the form
\begin{equation}
\varphi _{\alpha }(x^{i})=\frac{e^{i\mathbf{k}\cdot \mathbf{x}-i\omega t}}{%
\sqrt{2\omega (2\pi )^{D-1}}}f_{\lambda }(y),\quad \omega =\sqrt{%
k^{2}+\lambda ^{2}},\quad k=|\mathbf{k}|,  \label{eigfunc1}
\end{equation}%
where $\lambda $ is the separation constant. Below we will assume that $%
y\geqslant 0$. The corresponding formulae in the region $y<0$ are obtained
from the $Z_{2}$-symmetry of the model. Substituting eigenfunctions (\ref%
{eigfunc1}) into field equation (\ref{fieldeq}), for the function $%
f_{\lambda }(y)$ one obtains the equation
\begin{equation}
e^{-Du-w}\partial _{y}\left[ e^{Du-w}\partial _{y}f_{\lambda }\right]
-\left( m^{2}+\xi R-\lambda ^{2}e^{-2u}\right) f_{\lambda }=0.
\label{eqforfn}
\end{equation}%
For the exterior AdS geometry one has $u(y)=-k_{D}y$, $R=-D(D+1)k_{D}^{2}$
and the solution to equation (\ref{eqforfn}) is expressed in terms of
cylinder functions. The solution in the region $y<a$ even in $y$ we will
denote by $R(y,\lambda )$, $R(-y,\lambda )=R(y,\lambda )$. The parameter $%
\lambda $ enters in the radial equation in the form $\lambda ^{2}$ and this
solution can be chosen in such a way that $R(y,-\lambda )=\mathrm{const}%
\cdot R(y,\lambda )$. Now for the eigenfunctions one has%
\begin{equation}
f_{\lambda }(y)=\left\{
\begin{array}{ll}
R(y,\lambda ), & \mathrm{if}\;y<a, \\
e^{Dk_{D}y/2}\left[ A_{\nu }J_{\nu }(\lambda z)+B_{\nu }Y_{\nu }(\lambda z)%
\right] , & \mathrm{if}\;y>a,%
\end{array}%
\right.  \label{fl}
\end{equation}%
where $A_{\nu }$ and $B_{\nu }$ are integration constants, $J_{\nu }(x)$, $%
Y_{\nu }(x)$ are the Bessel and Neumann functions, and we use the notations
\begin{equation}
\nu =\sqrt{D^{2}/4-D(D+1)\xi +m^{2}/k_{D}^{2}},\;z=e^{k_{D}y}/k_{D}.
\label{nu}
\end{equation}%
For a conformally coupled massless scalar $\xi =(D-1)/(4D)$, $\nu =1/2$ and
the cylinder functions in Eq. (\ref{fl}) are expressed in terms of
elementary functions.

The radial function is continuous at $y=a$. In order to find the condition
for its derivative we note that the discontinuity of the function $u^{\prime
}(y)$ at $y=a$ leads to the delta function term $2D\left[ u^{\prime
}(a-)+k_{D}\right] \delta (y-a)$ in the Ricci scalar and, hence, in equation
(\ref{eqforfn}) for the radial eigenfunctions. For a non-minimally coupled
scalar field, due to the delta function term in the equation for the radial
eigenfunctions, these functions have a discontinuity in their slope at $y=a$%
. The corresponding jump condition is obtained by integrating the equation (%
\ref{eqforfn}) through the point $y=a$:%
\begin{equation}
f_{\lambda }^{\prime }(a+)-f_{\lambda }^{\prime }(a-)=\frac{16\pi GD\xi }{D-1%
}\tau _{0}^{0}f_{\lambda }(a).  \label{flderjump}
\end{equation}%
Now the coefficients in the formulae (\ref{fl}) for the exterior
eigenfunctions are determined by the continuity condition for the radial
eigenfunctions and by the jump condition for their radial derivative. From
these conditions for the radial part of the eigenfunctions in the region $%
y>a $ we find%
\begin{equation}
f_{\lambda }(y)=\frac{\pi }{2}e^{Dk_{D}(y-a)/2}R(a,\lambda )\left[ \bar{Y}%
_{\nu }(\lambda z_{a})J_{\nu }(\lambda z)-\bar{J}_{\nu }(\lambda
z_{a})Y_{\nu }(\lambda z)\right] .  \label{fl2}
\end{equation}%
where $z_{a}=e^{k_{D}a}/k_{D}$. Here and in what follows we use the notation%
\begin{equation}
\bar{F}(z)\equiv zF^{\prime }(z)+\left[ \frac{D}{2}-\frac{16\pi GD\xi }{%
(D-1)k_{D}}\tau _{0}^{0}-\frac{\partial _{y}R(y,\lambda )|_{y=a}}{%
k_{D}R(a,\lambda )}\right] F(z).  \label{barrednew}
\end{equation}%
Note that due to our choice of the function $R(y,\lambda )$, the logarithmic
derivative in formula (\ref{barrednew}) is an even function of $\lambda $.
From the orthonormalization condition for the radial eigenfunctions one
finds the relation%
\begin{equation}
R^{-2}(a,\lambda )=\frac{\pi ^{2}}{2}\frac{\bar{J}_{\nu }^{2}(\lambda z_{a})+%
\bar{Y}_{\nu }^{2}(\lambda z_{a})}{z_{a}^{D}k_{D}^{D-1}\lambda },
\label{normcoefRl}
\end{equation}%
which determines the normalization coefficient for the interior
eigenfunctions.

Substituting the eigenfunctions (\ref{eigfunc1}) into the mode sum (\ref%
{mfieldmodesum}), under the condition $z+z^{\prime }>2z_{a}+|t-t^{\prime }|$
the Wightman function can be presented in the form
\begin{eqnarray}
\fl \langle 0|\varphi (x)\varphi (x^{\prime })|0\rangle =\frac{1}{2}\langle
0_{S}|\varphi (x)\varphi (x^{\prime })|0_{S}\rangle -\frac{k_{D}^{D-1}}{%
(2\pi )^{D}}(zz^{\prime })^{\frac{D}{2}}\int d\mathbf{k}\,e^{i\mathbf{k}%
\cdot (\mathbf{x}-\mathbf{x}^{\prime })}\int_{k}^{\infty }d\lambda \lambda
\nonumber \\
\times \frac{\tilde{I}_{\nu }(\lambda z_{a})}{\tilde{K}_{\nu }(\lambda z_{a})%
}\frac{K_{\nu }(\lambda z)K_{\nu }(\lambda z^{\prime })}{\sqrt{\lambda
^{2}-k^{2}}}\cosh \!\left[ \sqrt{\lambda ^{2}-k^{2}}(t^{\prime }-t)\right] ,
\label{coreWF}
\end{eqnarray}%
where $\langle 0_{S}|\varphi (x)\varphi (x^{\prime })|0_{S}\rangle $ is the
positive frequency Wightman function for the AdS spacetime without
boundaries (see, for instance, \cite{Saha05}), and the second term on the
right is induced by the brane. Here and below the tilted notation for the
modified Bessel functions $I_{\nu }(x)$ and $K_{\nu }(x)$ is defined by the
formula%
\begin{equation}
\tilde{F}(x)\equiv xF^{\prime }(x)+\mathcal{R}(a,x)F(x),
\label{Barrednotmod}
\end{equation}%
with the notation%
\begin{equation}
\mathcal{R}(a,x)=\frac{D}{2}-\frac{16\pi GD\xi }{(D-1)k_{D}}\tau _{0}^{0}-%
\frac{\partial _{y}R(y,xe^{\pi i/2}/z_{a})|_{y=a}}{k_{D}R(a,xe^{\pi
i/2}/z_{a})}.  \label{newRlcal}
\end{equation}%
Quantum effects in free AdS spacetime are well investigated in literature
(see references given in \cite{Saha05}) and in the discussion below we will
be mainly concentrated on the effects induced by the brane.

As we see from (\ref{coreWF}), the information about the inner structure of
the brane is contained in the logarithmic derivative of the interior radial
function in formula (\ref{newRlcal}). In the RS 1-brane model with the brane
of zero thickness the brane induced part in the Wightman function is given
by a similar formula with the replacement \cite{Saha05}
\begin{equation}
\mathcal{R}(a,x)\rightarrow D/2-2D\xi -c/2k_{D},  \label{replaceR}
\end{equation}%
in the definition (\ref{Barrednotmod}) of the tilted notation. The parameter
$c$ is the brane mass term for a scalar field which is a phenomenological
parameter in the model with zero thickness brane. As we see, in the model
under consideration the effective brane mass term is determined by the core
structure. Note that in RS 2-brane model the mass terms on the branes
determine the 1-loop effective potential for the radion field and play an
important role in the stabilization of the interbrane distance.

\section{Vacuum expectation values outside the brane}

\label{sec:Outside}

Outside the brane the local geometry is the same as that for the AdS
spacetime and the renormalization procedure for the local characteristics of
the vacuum is the same as for the free AdS spacetime. By using the formula
for the Wightman function from the previous section, the VEV of the field
square in the exterior region is presented in the form%
\begin{equation}
\langle 0|\varphi ^{2}|0\rangle =\frac{1}{2}\langle 0_{S}|\varphi
^{2}|0_{S}\rangle +\langle \varphi ^{2}\rangle _{\mathrm{b}},
\label{phi2ext}
\end{equation}%
where $\langle 0_{S}|\varphi ^{2}|0_{S}\rangle $ is the VEV of the field
square in the free AdS spacetime. The part induced by the brane is obtained
from the second term on the right of formula (\ref{coreWF}) in the
coincidence limit:%
\begin{equation}
\langle \varphi ^{2}\rangle _{\mathrm{b}}=-\frac{k_{D}^{D-1}z^{D}}{(4\pi
)^{D/2}\Gamma \left( D/2\right) }\int_{0}^{\infty }dxx^{D-1}\,\frac{\tilde{I}%
_{\nu }(xz_{a})}{\tilde{K}_{\nu }(xz_{a})}K_{\nu }^{2}(xz).  \label{phi2cext}
\end{equation}%
The VEV\ of the field square in the free AdS spacetime is well investigated
in literature \cite{Burg85} and does not depend on the spacetime point,
which is a direct consequence of the maximal symmetry of the AdS bulk.

At large distances from the brane, $z\gg z_{a}$, we introduce a new
integration variable $y=xz$ and expand the integrand over $z_{a}/z$. By
using the formula for the integral involving the square of the MacDonald
function, to the leading order we obtain%
\begin{equation}
\fl \langle \varphi ^{2}\rangle _{\mathrm{b}}=-\frac{k_{D}^{D-1}(z_{a}/z)^{2%
\nu }}{2^{D+2\nu +1}\pi ^{(D-1)/2}}\frac{\mathcal{R}(a,0)+\nu }{\mathcal{R}%
(a,0)-\nu }\frac{\Gamma (D/2+\nu )\Gamma (D/2+2\nu )}{\nu \Gamma ^{2}(\nu
)\Gamma ((D+1)/2+\nu )}.  \label{phi2cextfar}
\end{equation}%
As we see, at large distances from the brane the brane induced part is
exponentially suppressed by the factor $\exp (-2\nu k_{D}y)$.

Having the Wightman function and the VEV\ for the field square, the VEV of
the EMT in the region $y>a$ can be evaluated by using the formula
\begin{eqnarray}
\langle 0|T_{ik}|0\rangle  &=&\lim_{x^{\prime }\rightarrow x}\partial
_{i}\partial _{k}^{\prime }\langle 0|\varphi (x)\varphi (x^{\prime
})|0\rangle   \nonumber \\
&&+\left[ \left( \xi -\frac{1}{4}\right) g_{ik}\nabla _{l}\nabla ^{l}-\xi
\nabla _{i}\nabla _{k}-\xi R_{ik}\right] \langle 0|\varphi ^{2}|0\rangle .
\label{mvevEMT}
\end{eqnarray}%
Note that on the left of this formula we have used the expression for the
EMT which differs from the standard one by the term which vanishes on the
solutions of the field equation (\ref{fieldeq}) (see Ref. \cite{SahaSurf}).
Similar to the Wightman function, the components of the vacuum EMT are
presented in the decomposed form%
\begin{equation}
\langle 0|T_{ik}|0\rangle =\frac{1}{2}\langle 0_{S}|T_{ik}|0_{S}\rangle
+\langle T_{ik}\rangle _{\mathrm{b}},  \label{Tikextdecomp}
\end{equation}%
where $\langle 0_{S}|T_{ik}|0_{S}\rangle $ is the vacuum EMT in
the free AdS spacetime and the part $\langle T_{ik}\rangle
_{\mathrm{b}}$ is induced by the brane. For a conformally coupled
massless scalar field and for even values $D$ the renormalized
free AdS part in the VEV of the EMT vanishes. For odd values of
$D$, this part is completely determined by the trace anomaly (see
\cite{Birrell}).

Substituting the expressions of the Wightman function and the VEV of the
field square into formula (\ref{mvevEMT}), for the part of the EMT induced
by the brane one obtains%
\begin{equation}
\langle T_{i}^{k}\rangle _{\mathrm{b}}=-\frac{k_{D}^{D+1}z^{D}\delta _{i}^{k}%
}{(4\pi )^{D/2}\Gamma \left( D/2\right) }\int_{0}^{\infty }dxx^{D-1}\,\frac{%
\tilde{I}_{\nu }(xz_{a})}{\tilde{K}_{\nu }(xz_{a})}F^{(i)}[K_{\nu }(xz)],
\label{Tikc}
\end{equation}%
where for a given function $g(v)$ we have introduced the notations
\begin{eqnarray}
F^{(i)}[g(v)] &=&\left( \frac{1}{2}-2\xi \right) \left[ v^{2}g^{\prime
2}(v)+\left( D+\frac{4\xi }{4\xi -1}\right) vg(v)g^{\prime }(v)\right.
\nonumber \\
&&+\left. \left( \nu ^{2}+v^{2}+\frac{2v^{2}}{D(4\xi -1)}\right) g^{2}(v)%
\right] ,  \label{Finew} \\
F^{(D)}[g(v)] &=&-\frac{v^{2}}{2}g^{\prime }{}^{2}(v)+\frac{D}{2}\left( 4\xi
-1\right) vg(v)g^{\prime }(v)  \nonumber \\
&&+\frac{1}{2}\left[ v^{2}+\nu ^{2}+2\xi D(D+1)-D^{2}/2\right] g^{2}(v),
\label{FDnew}
\end{eqnarray}%
with $i=0,1,\ldots ,D-1$. For a conformally coupled massless scalar field
one has $\nu =1/2$ and from formulae (\ref{Finew}), (\ref{FDnew}) it follows
that $F^{(i)}[K_{\nu }(x)]=F^{(D)}[K_{\nu }(x)]=0$. Hence, in this case the
brane induced parts in the VEVs of the EMT vanish. Note that for a
conformally coupled scalar and for even values $D$ the conformal anomaly is
absent and the free AdS part in the vacuum EMT vanishes as well.

For large distances from the brane, $z\gg z_{a}$, introducing a new
integration variable $y=xz$ we expand the integrand over $z_{a}/z$. To the
leading order this leads to the result%
\begin{equation}
\fl\langle T_{i}^{k}\rangle _{\mathrm{b}}=-\frac{2^{1-D-2\nu
}k_{D}^{D+1}\delta _{i}^{k}}{\pi ^{D/2}\Gamma \left( D/2\right) \nu \Gamma
^{2}(\nu )}\left( \frac{z_{a}}{z}\right) ^{2\nu }\frac{\mathcal{R}(a)+\nu }{%
\mathcal{R}(a)-\nu }\int_{0}^{\infty }dx\,x^{D+2\nu -1}F^{(i)}[K_{\nu }(x)].
\label{TikLargez}
\end{equation}%
The integrals in this formula may be evaluated by using the formulae from
\cite{Prud86}. Note that the free AdS parts in the VEVs of both field square
and the EMT do not depend on the spacetime point and, hence, at large
distances from the brane they dominate in the total VEVs. Noting that in the
limit of strong gravitational field in the region outside the brane,
corresponding to large values $k_{D}$, one has $z/z_{a}=e^{k_{D}(y-a)}\gg 1$%
, we see that formulae (\ref{phi2cextfar}), (\ref{TikLargez}) also describe
the asymptotic behavior of the brane induced VEVs in this limit. Hence, in
the limit of strong gravitational field, for the points not too close to the
brane boundary, the brane induced parts are exponentially suppressed. The
free AdS parts behave as $k_{D}^{D-1}$ and their contribution dominates for
strong gravitational fields.

\section{Model with flat spacetime inside the brane}

\label{sec:flowerpot}

\subsection{Exterior region}

\label{subsec:flpotExt}

As an application of the general results given above let us
consider a simple example assuming that the spacetime inside the
brane is flat. The corresponding models for the cosmic string and
global monopole cores were considered in \cite{Alle90} and are
known as flower-pot models. From the continuity condition on the
brane boundary it follows that in
coordinates $(x^{\mu },y)$ for the interior functions one has $u(y)=-k_{D}a$%
, $w(y)=0$. From the matching condition (\ref{matchcond2}) we find the
corresponding surface EMT with the non-zero components $\tau
_{i}^{k}=(D-1)k_{D}\delta _{i}^{k}/8\pi G$, $i=0,1,\ldots ,D-1$. The
corresponding surface energy density is positive. We consider the VEVs in
the exterior and interior regions separately.

For the model under consideration the interior radial eigenfunction with $%
Z_{2}$-symmetry, $R(-y,\lambda )=R(y,\lambda )$, has the form%
\begin{equation}
\fl R(y,\lambda )=\frac{2z_{a}^{D}k_{D}^{D-1}\lambda \cos (k_{y}y)}{\pi
^{2}\cos ^{2}(k_{y}a)\left[ \bar{J}_{\nu }^{2}(\lambda z_{a})+\bar{Y}_{\nu
}^{2}(\lambda z_{a})\right] },\;k_{y}^{2}=\lambda ^{2}e^{2k_{D}a}-m^{2}.
\label{Rlflow}
\end{equation}%
where the normalization coefficient is found from formula (\ref{normcoefRl})
and the barred notation is defined by%
\begin{equation}
\bar{F}(z)\equiv zF^{\prime }(z)+\left[ D/2-2\xi D+(k_{y}/k_{D})\tan (k_{y}a)%
\right] F(z).  \label{barredflow}
\end{equation}%
As a result, the parts in the Wightman function, in the VEVs of the field
square and the EMT induced by the brane are given by formulae (\ref{coreWF}%
), (\ref{phi2cext}) and (\ref{Tikc}) respectively, where the tilted
notations for the modified Bessel functions are defined by (\ref%
{Barrednotmod}) with the coefficient%
\begin{equation}
\mathcal{R}(a,x)=D/2-2\xi D-\sqrt{x^{2}+m^{2}/k_{D}^{2}}\tanh (ak_{D}\sqrt{%
x^{2}+m^{2}/k_{D}^{2}}).  \label{Rcalflowex}
\end{equation}%
Comparing (\ref{Rcalflowex}) with (\ref{replaceR}), we see that in the limit $%
a\rightarrow 0$ from the results of the model with flat interior spacetime
the corresponding formulae in the RS 1-brane model with a zero thickness
brane are obtained.

The VEVs of the field square and the EMT diverge on the boundary of the
brane. The leading term in the corresponding asymptotic expansion for the
field square is given by%
\begin{equation}
\langle \varphi ^{2}\rangle _{\mathrm{b}}\approx \frac{k_{D}A_{D}}{%
2^{D+5}\pi ^{(D+1)/2}}\frac{\Gamma ((D-1)/2)}{(D-2)(y-a)^{D-2}},\;D>2,
\label{phi2near}
\end{equation}%
with the notation%
\begin{equation}
A_{D}=4D-D^{2}+1+4\xi D(D-3)-4m^{2}/k_{D}^{2}.  \label{AD}
\end{equation}%
For $D\leqslant 2$ the VEV of the field square is finite on the core
boundary. Note that in the model with zero thickness brane located at $y=0$,
the corresponding VEVs near the brane behave as $y^{1-D}$.

For the asymptotic behavior of the EMT we have (no summation over $i$)%
\begin{equation}
\fl\langle T_{i}^{i}\rangle _{\mathrm{b}}\approx -\frac{k_{D}A_{D}(\xi -\xi
_{D})}{2^{D+4}\pi ^{(D+1)/2}}\frac{\Gamma ((D+1)/2)}{(y-a)^{D}},\;\langle
T_{D}^{D}\rangle _{\mathrm{b}}\approx \frac{Dk_{D}(y-a)}{D-1}\langle
T_{0}^{0}\rangle _{\mathrm{b}},  \label{T00near}
\end{equation}%
with $i=0,1,\ldots ,D-1$. For a conformally coupled field (no summation over
$i$)%
\begin{equation}
\fl\langle T_{i}^{i}\rangle _{\mathrm{b}}\approx \frac{D-3}{Dk_{D}(y-a)}%
\langle T_{D}^{D}\rangle _{\mathrm{b}}\approx \frac{m^{2}}{D}\langle \varphi
^{2}\rangle _{\mathrm{b}},\;i=0,1,\ldots ,D-1,\;D>3.  \label{Tiinearconf}
\end{equation}%
For $D=3$ the radial stress $\langle T_{D}^{D}\rangle
_{\mathrm{b}}$ diverges logarithmically. In the case $D=2$ the
corresponding VEVs are finite on the boundary of the brane. In the
limit $m/k_{D}\gg 1$ the brane-induced VEVs are suppressed by the
factor $\exp \left[ -2(m/k_{D})\ln (z/z_{a})\right] $.

\begin{figure}[tbph]
\begin{center}
\begin{tabular}{cc}
\epsfig{figure=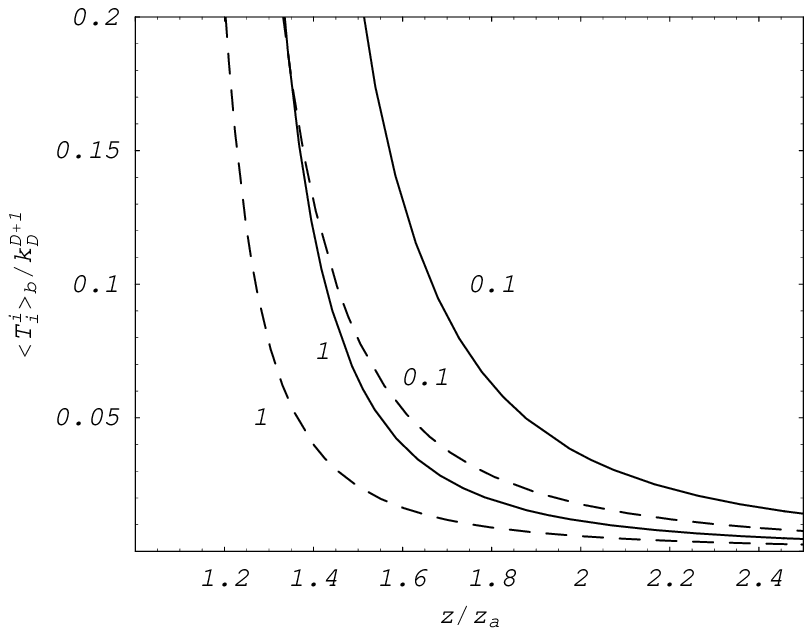,width=5.8cm,height=5cm} & \quad %
\epsfig{figure=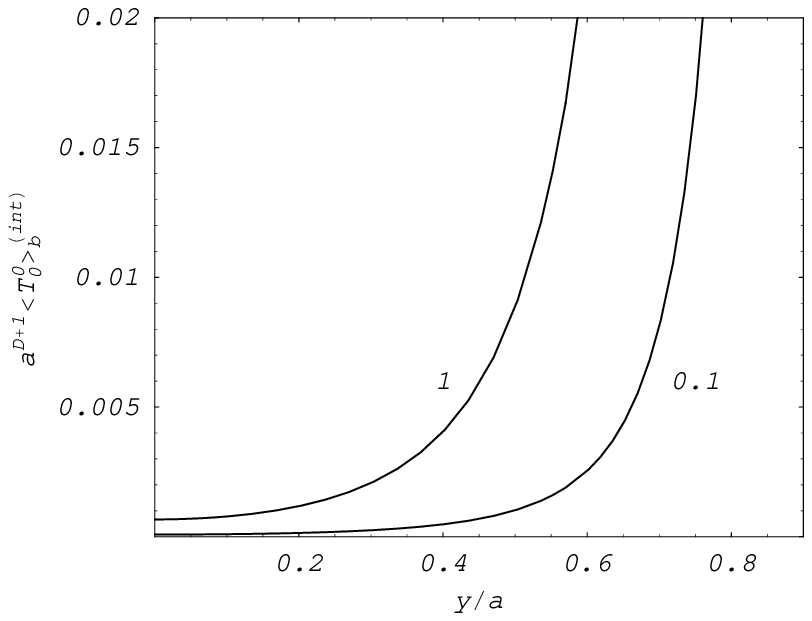,width=5.8cm,height=5cm}%
\end{tabular}%
\end{center}
\caption{On the left panel the part in the VEV of the energy density (full
curves) and radial stress (dashed curves), $k_{D}^{-D-1}\langle
T_{i}^{i}\rangle _{\mathrm{b}}$, $i=0,D$, induced by the brane are plotted
as functions of $z/z_{a}$ for a minimally coupled massless scalar field in $%
D=4$. On the right panel the corresponding energy density inside the brane, $%
a^{D+1}\langle T_{0}^{0}\rangle _{\mathrm{b}}^{\mathrm{(int)}}$, induced by
the AdS geometry in the exterior region is plotted as a function of $y/a$.
The numbers near the curves correspond to the values of $ak_{D}$. }
\label{figT00ex}
\end{figure}

On the left panel of figure \ref{figT00ex} we have plotted the dependence of
the brane induced parts in the VEVs of the energy density and radial stress
on $z/z_{a}$ for a minimally coupled massless scalar field ($\xi =0$) in the
case $D=4$. This parameter is related to the distance from the boundary of
the brane by the formula $z/z_{a}=\exp [k_{D}(y-a)]$. Recall that for a
conformally coupled massless scalar field the brane induced VEVs vanish.
Note that in $D=4$ the conformal anomaly is absent and for massless scalar
fields the VEV of the energy-momentum tensor in the free AdS spacetime is
zero.

\subsection{Interior region}

\label{subsec:flpotinter}

Now let us consider the vacuum polarization effects inside the brane for the
model with flat interior. Substituting eigenfunctions (\ref{Rlflow}) into
the mode sum formula, the corresponding Wightman function is presented in
the form
\begin{equation}
\langle 0|\varphi (x)\varphi (x^{\prime })|0\rangle =G_{0}(x,x^{\prime
})+G_{1}(x,x^{\prime }),  \label{G01}
\end{equation}%
where $G_{0}(x,x^{\prime })$ is the Wightman function in the Minkowski
spacetime orbifolded along the $y$-direction and%
\begin{eqnarray}
\fl G_{1}(x,x^{\prime }) =-\frac{(z_{a}k_{D})^{D}}{(2\pi )^{D}}\int d%
\mathbf{k\,}e^{i\mathbf{k}\cdot (\mathbf{x}-\mathbf{x}^{\prime
})}\int_{k}^{\infty }dx\,\frac{xC\{e^{-\varkappa (x)a},K_{\nu }(xz_{a})\}}{%
C\{\cosh (\varkappa (x)a),K_{\nu }(xz_{a})\}}  \nonumber \\
\times \frac{\cosh (\varkappa (x)y)\cosh (\varkappa (x)y^{\prime })}{%
\varkappa (x)\sqrt{x^{2}-k^{2}}}\cosh [\sqrt{x^{2}-k^{2}}(t-t^{\prime })].
\label{G1}
\end{eqnarray}%
In (\ref{G1}), $\varkappa (x)=\sqrt{x^{2}e^{2k_{D}a}+m^{2}}$, and we have
used the notation%
\begin{equation}
\fl C\left\{ f(u),g(v)\right\} =vf(u)g^{\prime }(v)+\left[ \left( D/2-2\xi
D\right) f(u)-(u/ak_{D})f^{\prime }(u)\right] g(v),  \label{Cfg}
\end{equation}%
with $g(v)=K_{\nu }(v)$ and $f(u)=e^{-u},\cosh u$ for the numerator and
denominator, respectively. The function $G_{0}(x,x^{\prime })$ differs by
the factor 1/2 from the Wightman function for a plate in the Minkowski
spacetime located at $y=0$ on which the field obeys the Neumann boundary
condition. The term $G_{1}(x,x^{\prime })$ is induced by the AdS geometry in
the region $y>a$. For a conformally coupled massless scalar field one has $%
\nu =1/2$ and by using definition (\ref{Cfg}) it can be explicitly checked
that $C\{e^{-u},K_{\nu }(u/ak_{D})\}=0$. Hence, in this case the part $%
G_{1}(x,x^{\prime })$ vanishes.

Now we turn to the evaluation of the renormalized VEV for the field square.
The renormalization corresponds to the omission of the part coming from the
Minkowskian Wightman function in (\ref{G01}). As a result the VEV is
presented in the form%
\begin{equation}
\langle \varphi ^{2}\rangle _{\mathrm{ren}}^{\mathrm{(int)}}=\langle \varphi
^{2}\rangle _{0,\mathrm{ren}}^{\mathrm{(int)}}+\langle \varphi ^{2}\rangle _{%
\mathrm{b}}^{\mathrm{(int)}}.  \label{phi2int}
\end{equation}%
Here the part $\langle \varphi ^{2}\rangle _{0,\mathrm{ren}}^{\mathrm{(int)}%
} $ is given by the formula%
\begin{equation}
\langle \varphi ^{2}\rangle _{0,\mathrm{ren}}^{\mathrm{(int)}}=\frac{m^{D-1}%
}{2(2\pi )^{(D+1)/2}}\frac{K_{(D-1)/2}\left( 2my\right) }{\left( 2my\right)
^{(D-1)/2}},  \label{phi2int0}
\end{equation}%
and the second term is obtained from (\ref{G1}) in the coincidence limit:%
\begin{equation}
\langle \varphi ^{2}\rangle _{\mathrm{b}}^{\mathrm{(int)}}=-\frac{(4\pi
)^{-D/2}}{\Gamma (D/2)}\int_{m}^{\infty }dx\,(x^{2}-m^{2})^{D/2-1}\cosh
^{2}(xy)U_{\nu }(x),  \label{phi2intb}
\end{equation}%
with the notation%
\begin{equation}
U_{\nu }(x)=\frac{C\{e^{-ax},K_{\nu }(\sqrt{x^{2}-m^{2}}/k_{D})\}}{C\{\cosh
(ax),K_{\nu }(\sqrt{x^{2}-m^{2}}/k_{D})\}}.  \label{Unu}
\end{equation}%
This part in the VEV of the field square is induced by the exterior AdS
geometry.

The integral on the right of formula (\ref{phi2intb}) is finite for $|y|<a$
and diverges on the boundary of the brane $|y|=a$. To the leading order,
near the boundary $y=a$ we find
\begin{equation}
\langle \varphi ^{2}\rangle _{\mathrm{b}}^{\mathrm{(int)}}\approx -\frac{%
k_{D}(\xi -\xi _{D})}{(4\pi )^{(D+1)/2}}\frac{D\Gamma ((D-1)/2)}{%
(D-2)(a-y)^{D-2}}.  \label{phi2intnear}
\end{equation}%
In the limit $am\gg 1$ the main contribution into the integral in formula (%
\ref{phi2intb}) comes from the lower limit and one finds%
\begin{equation}
\fl\langle \varphi ^{2}\rangle _{\mathrm{b}}^{\mathrm{(int)}}\approx -\frac{%
Bm^{D-1}\cosh ^{2}(my)}{(4\pi am)^{D/2}}e^{-2am},\;B\equiv \frac{D/2-2\xi
D+m/k_{D}-\nu }{D/2-2\xi D-m/k_{D}-\nu }.  \label{phi2amlarge}
\end{equation}%
As we could expect, in this limit the VEVs are exponentially suppressed.

As in the case of the field square, the VEV for the EMT is presented in the
form%
\begin{equation}
\langle T_{i}^{k}\rangle _{\mathrm{ren}}^{\mathrm{(int)}}=\langle
T_{i}^{k}\rangle _{0,\mathrm{ren}}^{\mathrm{(int)}}+\langle T_{i}^{k}\rangle
_{\mathrm{b}}^{\mathrm{(int)}},  \label{Tikint}
\end{equation}%
where $\langle T_{ik}\rangle _{0,\mathrm{ren}}^{\mathrm{(int)}}$ is the
vacuum EMT in the Minkowski spacetime orbifolded along the $y$-direction and
the presence of the part $\langle T_{ik}\rangle _{\mathrm{b}}^{\mathrm{(int)}%
}$ is related to that the geometry in the region $y>a$ is AdS. For the first
part one has
\begin{equation}
\fl\langle T_{i}^{k}\rangle _{0,\mathrm{ren}}^{\mathrm{(int)}}=\frac{%
m^{D+1}\delta _{i}^{k}}{(4\pi my)^{\frac{D+1}{2}}}\left[ K_{\frac{D+1}{2}%
}\left( 2my\right) \left( 2\xi -1\right) +\left( 1-4\xi \right) myK_{\frac{%
D+3}{2}}\left( 2my\right) \right] ,  \label{Tiiint0n}
\end{equation}%
with $\;i=0,1,\ldots ,D-1$ and $\langle T_{D}^{D}\rangle _{0,\mathrm{ren}}^{%
\mathrm{(int)}}=0$. For the second term on the right of (\ref{Tikint}) we
find (no summation over $i$)%
\begin{eqnarray}
\langle T_{i}^{i}\rangle _{\mathrm{b}}^{\mathrm{(int)}} &=&\frac{(4\pi
)^{-D/2}}{\Gamma (D/2)}\int_{m}^{\infty }dx(x^{2}-m^{2})^{D/2}U_{\nu }(x)
\nonumber \\
&&\times \left\{ \frac{1}{D}\cosh ^{2}(xy)+\frac{\left( 4\xi -1\right) x^{2}%
}{x^{2}-m^{2}}\left[ \cosh ^{2}(xy)-\frac{1}{2}\right] \right\} ,
\label{T00intb} \\
\langle T_{D}^{D}\rangle _{\mathrm{b}}^{\mathrm{(int)}} &=&-\frac{(4\pi
)^{-D/2}}{2\Gamma (D/2)}\int_{m}^{\infty
}dx\,(x^{2}-m^{2})^{D/2-1}x^{2}U_{\nu }(x),  \label{TDDintb}
\end{eqnarray}%
with $i=0,1,\ldots ,D-1$, and the function $U_{\nu }(x)$ is defined by
formula (\ref{Unu}). Note that the radial stress inside the brane does not
depend on spacetime point. This result could be also obtained directly from
the continuity equation. For a conformally coupled massless scalar we have $%
U_{\nu }(x)=0$ and, hence, the parts in the VEV of the EMT given
by (\ref{T00intb}),(\ref{TDDintb}) vanish. In this case the part
$\langle T_{i}^{i}\rangle _{0,\mathrm{ren}}^{\mathrm{(int)}}$
vanishes as well.

For the VEV of the EMT near the brane core we find (no summation over $i$)%
\begin{equation}
\fl\langle T_{i}^{i}\rangle _{\mathrm{b}}^{\mathrm{(int)}}\approx \frac{D-1}{%
Dk_{D}(y-a)}\langle T_{D}^{D}\rangle _{\mathrm{b}}^{\mathrm{(int)}}\approx
\frac{Dk_{D}(\xi -\xi _{D})^{2}}{2^{D}\pi ^{(D+1)/2}}\frac{\Gamma ((D+1)/2)}{%
(a-y)^{D}},  \label{Tiiintnear}
\end{equation}%
with $i=0,1,\ldots ,D-1$. For a conformally coupled scalar field the
corresponding asymptotic behavior is given by formulae (\ref{Tiinearconf}).
In the limit $am\gg 1$ to the leading order one has (no summation over $i$)%
\begin{equation}
\fl\langle T_{D}^{D}\rangle _{\mathrm{b}}^{\mathrm{(int)}}\approx -\frac{%
Bm^{D+1}e^{-2am}}{2(4\pi am)^{D/2}},\;\langle T_{i}^{i}\rangle _{\mathrm{b}%
}^{\mathrm{(int)}}\approx (1-4\xi )\left[ 2\cosh ^{2}(my)-1\right] \langle
T_{D}^{D}\rangle _{\mathrm{b}}^{\mathrm{(int)}}.  \label{TDDlargeam}
\end{equation}

On the right panel of figure \ref{figT00ex} we have plotted the dependence
of the part in the VEV of the energy density induced by the exterior AdS
geometry in the region inside the brane as a function of $y/a$ for a
minimally coupled massless scalar field in the case $D=4$. The corresponding
radial stress does not depend on $y$ and $\langle T_{D}^{D}\rangle _{\mathrm{%
b}}^{\mathrm{(int)}}\approx 0.00134/a^{5}$ for $ak_{D}=1$ and $\langle
T_{D}^{D}\rangle _{\mathrm{b}}^{\mathrm{(int)}}\approx 0.000173/a^{5}$ for $%
ak_{D}=0.1$. We recall that for a conformally coupled massless scalar field
the corresponding VEVs vanish for both field square and the EMT. The
perpendicular interior vacuum force acting per unit surface of the brane
boundary is determined by $-\langle T_{D}^{D}\rangle _{\mathrm{b}}^{\mathrm{%
(int)}}$. For minimally and conformally coupled scalars these forces tend to
decrease the brane thickness.

\section{Conclusion}

\label{sec:conc}

We have considered the one-loop vacuum effects for a massive scalar field
induced by a $Z_{2}$-symmetric thick brane on the $(D+1)$-dimensional AdS
bulk. Among the most important characteristics of the vacuum, which carry
information about the internal structure of the brane, are the VEVs for the
field square and the EMT. In order to obtain these expectation values we
first construct the Wightman function. In the region outside the brane this
function is presented as a sum of two distinct contributions. The first one
corresponds to the Wightman function in the free AdS geometry and the second
one is induced by the brane. The latter is given by formula (\ref{coreWF}),
where the tilted notation is defined by formula (\ref{Barrednotmod}) with
the coefficient from (\ref{newRlcal}). This coefficient is determined by the
radial part of the interior eigenfunctions and describes the influence of
the core properties on the vacuum characteristics in the exterior region.

In section \ref{sec:Outside} we have investigated the influence of the
non-trivial internal structure of the brane on the VEVs of the field square
and the EMT. The parts in these VEVs induced by the brane are directly
obtained from the corresponding part of the Wightman function. These parts
are given by formulae (\ref{phi2cext}) and (\ref{Tikc}) for the field square
and the EMT respectively. For a conformally coupled massless scalar field
the corresponding EMT vanishes. The parts in the VEVs of the field square
and EMT induced by the brane diverge on the boundary of the brane. At large
distances from the brane the brane induced VEVs are suppressed by the factor
$e^{-2\nu k_{D}y}$. In the limit of strong gravitational fields
corresponding to large values of the AdS energy scale $k_{D}$, for points
not too close to the brane the parts in the VEVs induced by the brane behave
as $k_{D}^{D\pm 1}e^{-2\nu k_{D}(y-a)}$ with upper/lower sign corresponding
to the EMT/field square. In this case the relative contribution of the brane
induced effects are exponentially suppressed with respect to the free AdS
part.

As an application of the general results, in section \ref{sec:flowerpot} we
have considered a simple model with flat spacetime in the region inside the
brane. The brane induced parts of the exterior VEVs in this model are
obtained from the general results by taking the function in the coefficient
of the tilted notation from Eq. (\ref{Rcalflowex}). We have also
investigated the vacuum densities inside the brane. Though the spacetime
geometry inside the brane is Monkowskian, the AdS geometry of the exterior
region induces vacuum polarization effects in this region as well. In order
to find the corresponding renormalized VEVs of the field square and the EMT
we have presented the Wightman function in the interior region in decomposed
form (\ref{G01}). In this representation the first term on the right is the
Wightman function in the Minkowski spacetime orbifolded along the direction
perpendicular to the brane and the second one is induced by the AdS geometry
in the exterior region. The corresponding parts in the VEVs of the field
square and the EMT are given by formulae (\ref{phi2intb}), (\ref{T00intb}), (%
\ref{TDDintb}). For a massless conformally coupled scalar field these parts
vanish. In the general case of the curvature coupling parameter, the
corresponding radial stress is uniform inside the brane and determines the
interior vacuum forces acting on the boundary of the brane. For both
minimally and conformally coupled scalar fields these forces tend to
decrease the thickness of the brane. When the brane thickness tends to zero,
from the formulae of the model with flat interior the corresponding results
in the RS 1-brane model are obtained.

\section*{Acknowledgments}

AAS is grateful to the organizers of QFEXT07 for financial support. The work
was supported by the Armenian Ministry of Education and Science Grant No.
0124.

\end{document}